\documentclass[journal]{IEEEtran}
\usepackage{amssymb}
\usepackage{caption}
\usepackage{graphicx}
\usepackage{xcolor}
\usepackage{subfig}
\usepackage{algorithm}
\usepackage{algpseudocode}
\usepackage{setspace}
\usepackage{float}
\usepackage{cite}
\usepackage{comment}
\usepackage{xcolor}
\usepackage{makecell}
\algnewcommand{\Initialize}[1]{%
  \State \textbf{Initialize:}
}
\algnewcommand{\Output}[1]{%
  \State \textbf{Output:}
}

\usepackage{amsmath}

\usepackage{mathtools}

\ifCLASSINFOpdf

\else

\fi
\begin{document}
\title{Computational Ghost Imaging with Low-Density Parity-Check Code}

\author{Shuang~Liu, Yunkai~Hu, Jinquan~Qi, Shensheng~Han, 
  and Zihuai~Lin}
  





\maketitle

\begin{abstract}
Ghost imaging (GI) is a high-resolution imaging technology that has been a subject of interest to many fields in the past 20 years. Most GI researchers focus on the reconstruction of signal under-sampling, nevertheless, how to use information redundancy to improve the result's belief in a complex environment has hardly been studied.
Motivated by this, we propose a computational GI system based on the low-density parity-check (LDPC) coded radiation field by exploiting the signal redundancy. The non-ideal factors generated within the imaging process can be eliminated by setting up the matching fading channel model. We have derived the analytical lower bound on the bit error rate for the proposed LDPC-coded GI system. The effectiveness and performance of the LDPC-coded GI system are further validated through numerical and experiment results. 
\end{abstract}

\begin{IEEEkeywords}
Channel coding, computational ghost imaging, linear block code, imaging system, information theory, signal detection   
\end{IEEEkeywords}

\IEEEpeerreviewmaketitle


\section{INTRODUCTION}
\label{intro}

\begin{figure}[t!]
\centering
\includegraphics[width=9.5cm]{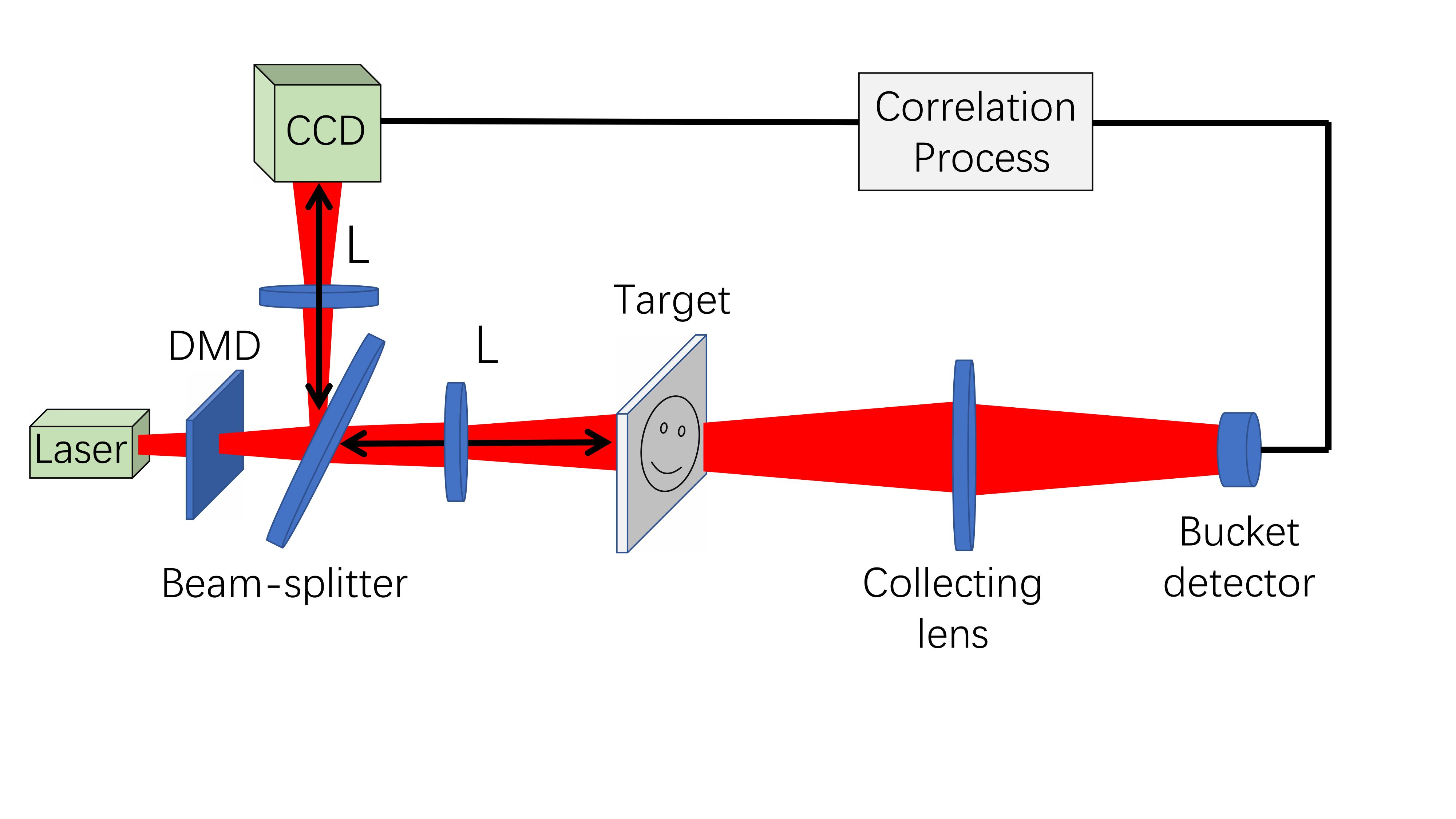}
\caption{The system structure of an optical GI. This figure shows the standard pseudo-thermal two-detector setup. The light field from the laser is mapped on the DMD, and the spatially modulated light field is divided into two paths; The pseudo-thermal field received by the CCD is recorded as the reference path, and the 1-D signal received by the bucket detector is recorded as the transmitting path.}
\label{OGI}
\end{figure}

\IEEEPARstart{G}{host} imaging (GI) is a non-local imaging technique that reconstructs an unknown image by computing the correlation between a speckle field, which is generated by a digital micro-mirror device (DMD) or a ground glass, that passes through an object and the one-dimensional (1-D) signal received by a bucket detector, in which the bucket detector can collect the energy in the aperture. The spatial and temporal coded patterns modulated on the transmitting radiation field are formed specifically\cite{OGI1,OGI2,OGI3,OGI4,OGI5,OGI6}. The optical GI system is set as an example shown in Fig. \ref{OGI}. When the transmitting speckle field reaches the target surface, the speckle field and the target reflectance are spatial and temporal modulated. In essence, GI is to form an irradiated object with spatial fluctuation by modulating the radiation source, and the image of the object is reconstructed by correlating the radiation source with the intensity of the received signal from a bucket detector. GI has been applied in various waveband systems, such as X-ray\cite{Xray1,Xray2,Xray3}, microwave\cite{MGI2,ZH1,ZH2,ZH3,GFKD}, terahertz\cite{Thz1,Thz2} and optical waveband\cite{OP1,OP2,OP3}. Therefore, we use the radiation field to represent the radiation source of different bands in this work.\\

\begin{figure}[t!]
\centering
\includegraphics[width=8.8cm,height = 5.4cm]{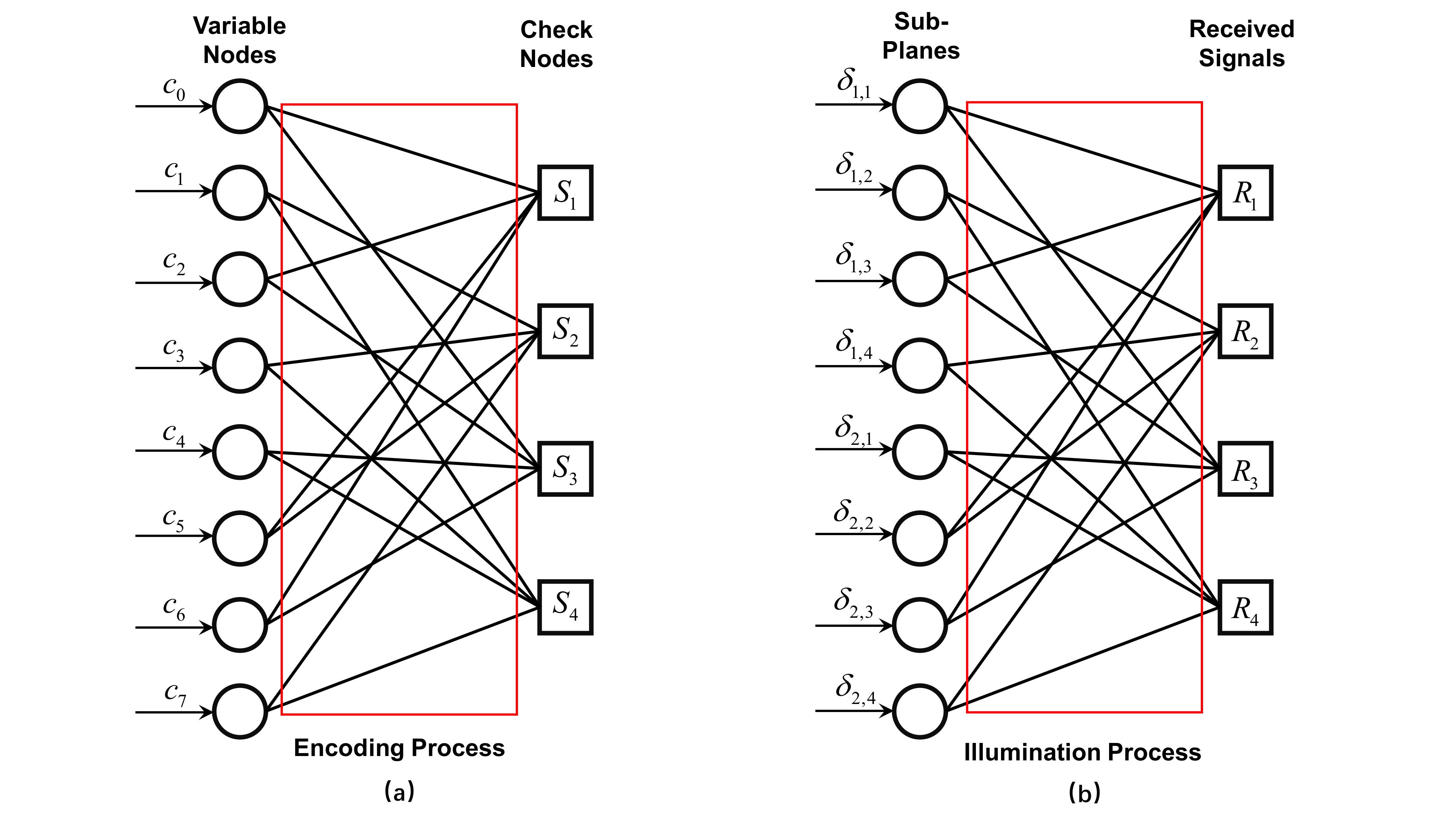}
\caption{The factor graph representation of (a) 8 channels LDPC-code encoding process; and (b) conventional GI for 8 pixels ($N=2, P=4$) resolution. Note that the LDPC-code encoding process shown in (a) does not contain the internal structure of information symbols. Similarly, the GI process represented in (b) does not detail the spatial relationship between different reflectivity.}
\label{factor graph}
\end{figure}

\begin{figure*}[h!]
\centering
\includegraphics[width=16cm,height = 7cm]{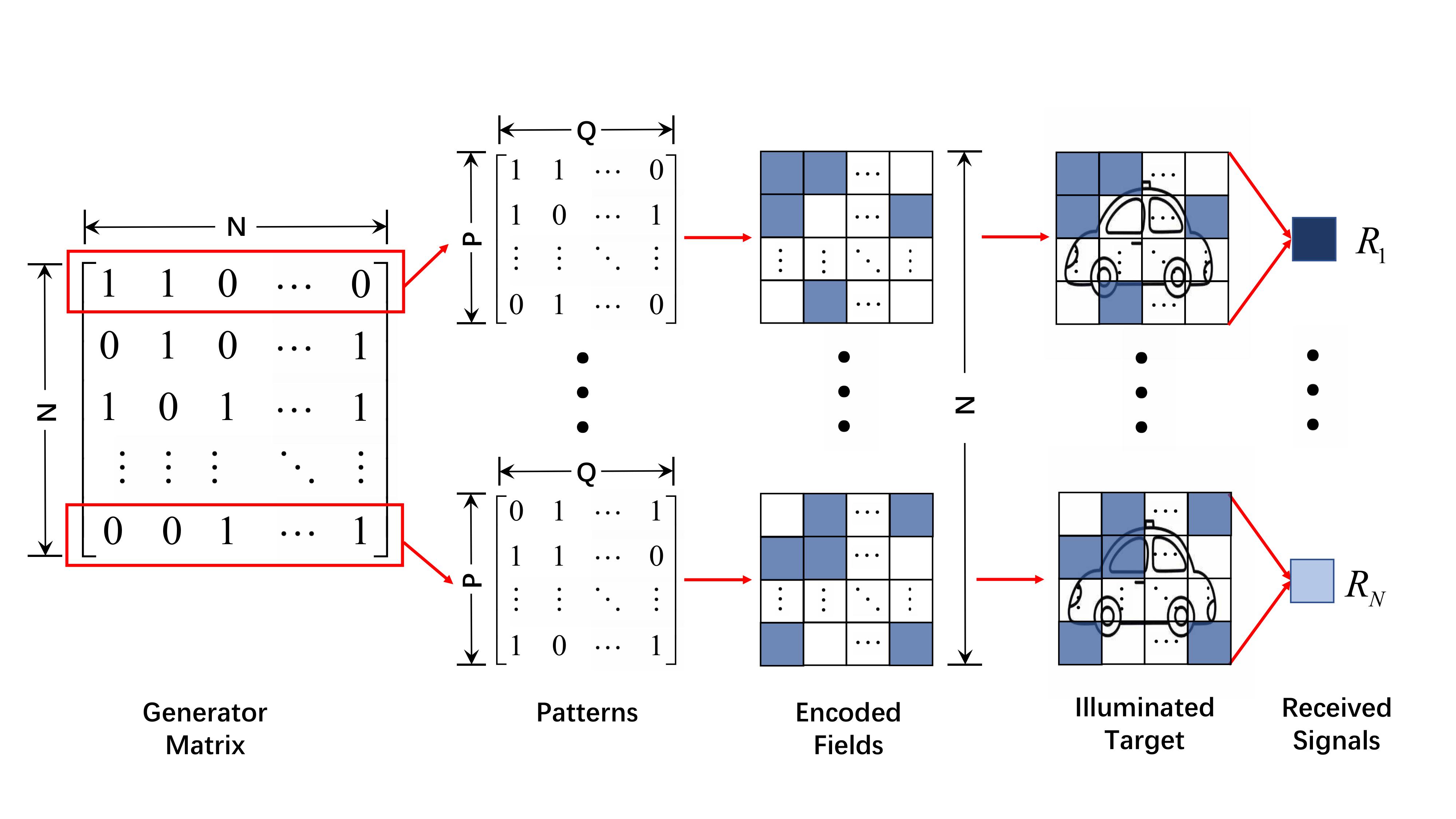}
\caption{The procedures of proposed GI method using LDPC code encoded fields.}
\label{whole process}
\end{figure*}

Generally, the actual sensing systems are always affected by the interference like background noise, scattering medium, atmosphere turbulence and etc. Repeated measurement, which is often used by the conventional GI system, is a common method to overcome the interference. A great number of measurements can let the noise interference gradually approach its mean value. However, this method needs a long measuring time whose efficiency is low. Meanwhile, the error control coding in wireless communication is widely used to protect the information from the interference. For example, LDPC\cite{LDGM}, as a class of error control coding, can monitor the information transferring errors in telecommunication to protect the information accuracy. Motivated by this, we propose the channel-coded GI with LDPC codes, which can protect the target information from the distortion occurred in the imaging procession. The detecting radiation field can be encoded by shaping LDPC from one-dimensional form to two dimensions. Thus once the main interference factor is accurately modeled, the target information can be reconstructed by the belief propagation (BP) decoding algorithm.

The image of the target can be reconstructed by iteratively updating the probability domain information of the information bits. Differing from statistical decoding methods given by conventional random encoding, our proposed LDPC-coded GI can be theoretically analyzed, in which we have derived the analytical lower bound on the decoding bit error rate (BER) of the LDPC-coded GI system.\\



 The rest of this paper is organized as follows: Section \ref{2} presents the generation process of the encoded speckle field in GI. Section \ref{4} gives the analytical lower bound of the LDPC-coded GI imaging performance. Section \ref{5} provides the numerical results to compare the performance of our proposed LDPC-coded GI system with conventional binary-coded GI systems, and verify the correctness of the theoretical analysis of imaging performance. Section \ref{6} verifies the theoretical feasibility of our proposed LDPC-coded GI system through a series of experiments. Finally, the conclusion is drawn in Section \ref{7}.

\begin{table}[h!t]
\centering
 \caption{The BP decoding process.}
\begin{tabular}{|c|c|}
  \hline
  Number & Steps  \\
  \hline
  First & Initialization  \\
                                          
  \hline
  Second & \makecell*[c]{Updating variable nodes information } \\
  \hline
  Third & \makecell*[c]{Updating check nodes information } \\
  \hline
  Forth & Updating the hard decision \\
  \hline
  Fifth & Terminating the iteration\\
  
  \hline
  
  \end{tabular}
 \label{BP_table}
\end{table}

\section{IMAGING WITH LDPC CODING ENCODED FIELD}
\label{2}

In this section, the generation of the encoded field is introduced. As illustrated in Fig. \ref{factor graph}, no matter by which physical means an imaging system is constructed, it can be considered as an information transfer process. Therefore, we can find some commons between communication systems\cite{a33} and imaging systems. A classic communication system consists of three parts: the information source to be transmitted, the information transmitting process, and the data sink. Theoretically, the imaging system has three parts similar to the communication system: the imaging information awaits to be acquired, the detecting process, and the receiver. The information transfer process is the major difference between these two systems. 

Let us take an example, in a communication system, a $[N, M, \Omega_{\omega(\textbf{p})}]$ LDPC code encodes a block of $M$ information bits into a $N$ bits codeword with a sparse generator matrix, $\textbf{G}=[\textbf{I} | \textbf{P}]$, where $\textbf{I}$ is an identity matrix of size $M$ and $\textbf{P}$ is a randomly generated parity matrix of size $K \times (N-K)$. Therefore, the code rate is $R_{c}=K/N$. The Hamming weight $D$ ($1 \le D \le K$) at each column of $\textbf{P}$ are indicated by the degree distribution $\Omega (x)$, where $\Omega (x)=\sum_D^K \Omega_D x^{D}$.

the information is encoded by the LDPC code and transmitted over 1-D communication channels. The 2-bit information can be decoded with the BP decoding algorithm. As the same as the former, the information of imaging systems is in the two-dimensional (2-D) imaging plane.\\

Without loss of generality, we consider the investigation area at one time instant to be a $2$-D plane $D$, which is shown in Fig. \ref{whole process}. If needed, an extension to $3$-D space can be readily accommodated. The imaging plane $D$ can be divided equally into $P$ rows and $Q$ columns sub-planes.
The number of illumination cycles $N$ satisfies the condition $N=P \times Q$.


After $N$ illuminations, the illumination matrix on the surface of the target plane can be denoted as
\begin{equation}
\mathbf{A}=\left[\begin{array}{cccc}
A_{1,1}^{(1)} & A_{2,1}^{(1)} & \cdots & A_{P, Q}^{(1)} \\
A_{1,1}^{(2)} & A_{2,1}^{(2)} & \cdots & A_{P, Q}^{(2)} \\
\vdots & \vdots & \ddots & \vdots \\
A_{1,1}^{(N)} & A_{2,1}^{(N)} & \cdots & A_{P, Q}^{(N)}
\end{array}\right],
\end{equation}
where $A_{p, q}^{(n)}$ represents the speckle point form located at $p$ row and $q$ column during the $n$-th illumination. Due to the bucket detector, the received signal is the incoherent superposition of reflected signals from scattering points on the imaging plane. The received signal from the bucket detector after $N$ illuminations can be written as
\begin{equation}
\mathbf{R}=\left[R_{1}, R_{2}, \cdots, R_{N}\right]^{T}.
\end{equation}

Then we rearrange the imaging plane row by row to form a vector of the scattering coefficient as
\begin{equation}
\boldsymbol\delta=\left[\delta_{1,1}, \delta_{2,1}, \cdots, \delta_{P, Q}\right]^{T}.
\end{equation}

Therefore, the whole detection can be represented by the linear model as
\begin{equation}
\mathbf{R}=\mathbf{A}\boldsymbol\delta.
\end{equation}

A unique solution of the target reflectance $\boldsymbol\delta$ will be found if the matrix $\mathbf{A}$ is of full rank. The above construction algorithm is an ideal case. However, In the real situation, noise is always present at the receiver.  As shown in Fig. \ref{shiyi}, the transmitting optical field information conforms to the additive circular symmetric Gaussian distribution. When the imaging is incoherent, the additive circular symmetric Gaussian distribution degenerates into the Rayleigh fading distribution, denoted by $\mathbf{h}=[h_{1}, \cdots, h_{N}]^{T}$, and $\mathbf{h}$ are i.i.d $\mathcal{CN} (0, 1)$. The other is the Additive White Gaussian Noise (AWGN) denoted by $\mathbf{n}$, where
$\mathbf{n}$ are i.i.d $\mathcal{CN} (0, N_0)$. Thus, the updated received signal can be written as
\begin{equation}
\mathbf{R}=\mathbf{A} \cdot ( \mathbf{h} \boldsymbol\delta) + \mathbf{n}.
\label{eqr1}
\end{equation}

Under the noisy environment, the essence of the reconstruction algorithm turns into the problem of minimizing the function $||\mathbf{R- \mathbf{A} \cdot ( \mathbf{h} \boldsymbol\delta)}||$. Some solutions can be found in the literature, such as the Gradient Projection (GP) algorithm\cite{a39}. Also, the target image can be reconstructed by the BP algorithm \cite{BP}, whose steps are summarized in Table. \ref{BP_table}. How to eliminate the interference of these non-ideal factors by channel coding becomes the key to this imaging model.

\begin{figure*}[!htb]
\begin{align}
P_b=\frac{1}{2}\Bigg(\bigg(1-\sqrt{\frac{\gamma}{1+\gamma}}\bigg)+\sum_{j=0}^{N-K}  \binom{N-K}{j} \aleph_{1}^j (1-\aleph_{1})^{N-K-j} \text{erfc}\bigg(\sqrt{\frac{\sum_{j \in \phi}{E_s^j}}{R_cN_0}}\bigg)\Bigg),
\label{lo1}
\end{align} 
where
\begin{align}
\gamma=\frac{E_s}{N_0}.\nonumber
\end{align}
\hrulefill
\end{figure*}

\section{IMAGING PERFORMANCE Analysis OF LDPC CODING ASSISTED GHOST IMAGING}
\label{4}
In this section, the theoretical BER lower bound of the LDPC-coded GI imaging system will be given in the probability domain.

Assuming that the light field source space is entirely incoherent and the irradiation is uniform, whose value is $E_s$. Meanwhile, the transverse dimension of space is large enough compared with the target dimension. For a binary-valued target, an error would occur for the $i$-th pixel in the imaging plane when the value in this pixel satisfies
\begin{equation}
    \left | \hat{\boldsymbol{\delta}}_i - \left \langle  \hat{\boldsymbol{\delta}}_i \right\rangle   \right | > 0.5E_s.
\end{equation}

When there is no object in the field of view (so-called `pure noise transmission'), the target function becomes $\boldsymbol{\delta} = \boldsymbol{0}^{T}$.

From Eq.~(\ref{eqr1}), we can obtain the theoretical BER lower bound of the LDPC-coded GI system as
\begin{align}
P_b=P_{\text{ray}}+P_e.  
\label{hu0}
\end{align}

The first term $P_{\text{ray}}$ in the summand of Eq.~(\ref{hu0}) is the BER of the Rayleigh fading from the target to the optical lens. The second term $P_e$ is the decoding error probability of the signal that is encoded at the DMD and received at the bucket detector. The decoding error probability is lower bound by the probability that one pixel in the final image cannot be recovered correctly. 

Based on the computational method of the BP decoding algorithm, the probability that there exists one pixel that cannot be recovered correctly can be expressed as 
\begin{equation}
\begin{aligned}
&P_e=\operatorname{Pr}\big(\hat{\boldsymbol{\delta}}=\textbf{e}: \textbf{e}\mathbf{A}  \neq \boldsymbol{0}| \omega(\textbf{e}) =1\big),
\end{aligned}
\label{8}
\end{equation}
where $\textbf{e}$ is the recovered 1-D image sequence with errors, and $\omega(\textbf{e})$ is the hamming weight whose value stands for the number of error pixels. 

The probability $\Pr\big(\textbf{e}\mathbf{A}  \neq \mathbf{0} | \omega(\textbf{e}) =1\big)$ in Eq.(\ref{8}) represents the probability that the recovered image sequence multiplying with the generator matrix $\textbf{A}$ is not an all-zero, in which $\textbf{A}=[\textbf{I}|\textbf{P}]$. In this case, the recovered image sequence is identical to the first $K$ bits of the decoded codeword, and the rear $N-K$ bits contain non-zeros if $\textbf{e}\textbf{P}\neq\mathbf{0}$.

Given the recovered image sequence has an error pixel $\omega(\textbf{e})=1$ and the column vector $\textbf{p}$ of the parity  matrix $\textbf{P}$ has weight $\omega(\textbf{p})$, the probability that the recovered image sequence multiplying with the column vector of the parity matrix is not a zero can be expressed as
\begin{align}
\aleph_{1}&= \Pr\big(\textbf{e}\otimes\mathbf{p}  \neq \mathbf{0} | \omega(\textbf{e}) =1\big) \nonumber\\
&=\sum_{\omega(\textbf{p})} \Omega_{\omega(\textbf{p})} \Pr\big( \textbf{e}\otimes\textbf{p} \neq \mathbf{0} | \omega(\textbf{p}), \omega(\textbf{e})=1\big),
\label{hu1}
\end{align}
where $\otimes$ represents vector multiplication over GF(2), and the weight $\omega(\textbf{p})$ of the column vector $\textbf{p}$ is randomly and independently generated by the degree distribution $\Omega_{\omega(\textbf{p})}$.

Let us denote by $\kappa$ the position of the error pixel in the recovered image sequence, and $\textbf{p}(\kappa)$ be the corresponding subvector of $\textbf{p}$. Then the probability $\Pr\big( \textbf{e}\otimes\textbf{p} \neq \mathbf{0} | \omega(\textbf{p}), \omega(\textbf{e})=1\big)$ in Eq.~(\ref{hu1}) is equal to the probability that $\textbf{p}(\kappa)$ has an one at the exact position of $\textbf{e}$. Therefore, we have
\begin{align}
 \Upsilon_1^{\omega(\textbf{p})}& = \Pr\big( \textbf{e} \otimes \textbf{p} \neq \mathbf{0} | \omega(\textbf{p}), \omega(\textbf{e})=1\big) \nonumber\\
 &=\cfrac{ \binom{K-1}{\omega(\textbf{p})-1} }{\binom{K}{\omega(\textbf{p})}} . 
 \label{ld2}
\end{align}

By adding up all the cases of weight
$\omega(\textbf{p})$ associated with degree distribution $\Omega_{\omega(\textbf{p})}$, the probability that the weight-1 error sequence multiplies with the column vector $\textbf{p}$ is not a zero can be expressed as
\begin{align}
\aleph_{1}&=\Pr\big(\textbf{e}\otimes\mathbf{p}  \neq \mathbf{0} | \omega(\textbf{e}) =1\big) \nonumber\\
&=\sum_{\omega(\textbf{p})} \Omega_{\omega(\textbf{p})} \Upsilon_1^{\omega(\textbf{p})}.
\label{de4}
\end{align}

The probability $\Pr\big(\textbf{e}\mathbf{A}  \neq \mathbf{0} | \omega(\textbf{e}) =1\big)$ in Eq.(\ref{8}) is conditioned on the probability $\Pr(\textbf{R}' \rightarrow \textbf{R})$, which gives the weight-1 error sequence. Since the process from the DMD to the bucket detector follows the AWGN channel, we have
\begin{align}
\Pr(\textbf{R}' \rightarrow \textbf{R})=\frac{1}{2} \text{erfc}\bigg(\sqrt{\frac{\sum_{j \in \phi}{E_s^j}}{R_cN_0}}\bigg),
\label{hu2}
\end{align}
where $E_s^j$ denotes the transmitted symbol energy, and $R_c$ is the code rate. ${j \in \phi}$ is the number of incorrect symbols in the received codeword, giving $\omega(\textbf{e}) =1$.

By combining Eq.~(\ref{8}), Eq.~(\ref{de4}), and Eq.~(\ref{hu2}), the probability that there exists one pixel that cannot be recovered correctly can be finally obtained by
\begin{align}
P_e=&\frac{1}{2}\sum_{j=0}^{N-K}  \binom{N-K}{j} \aleph_{1}^j (1-\aleph_{1})^{N-K-j} \nonumber\\ &\cdot \text{erfc}\bigg(\sqrt{\frac{\sum_{j \in \phi}{E_s^j}}{R_cN_0}}\bigg).
\end{align}

Finally, the analytical BER lower bound of the LDPC-coded GI system can be summarized in Eq.~(\ref{lo1}).

\section{NUMERICAL RESULTS}
\label{5}
\begin{figure}[!t]
\centering
\includegraphics[width=8cm,height=7cm]{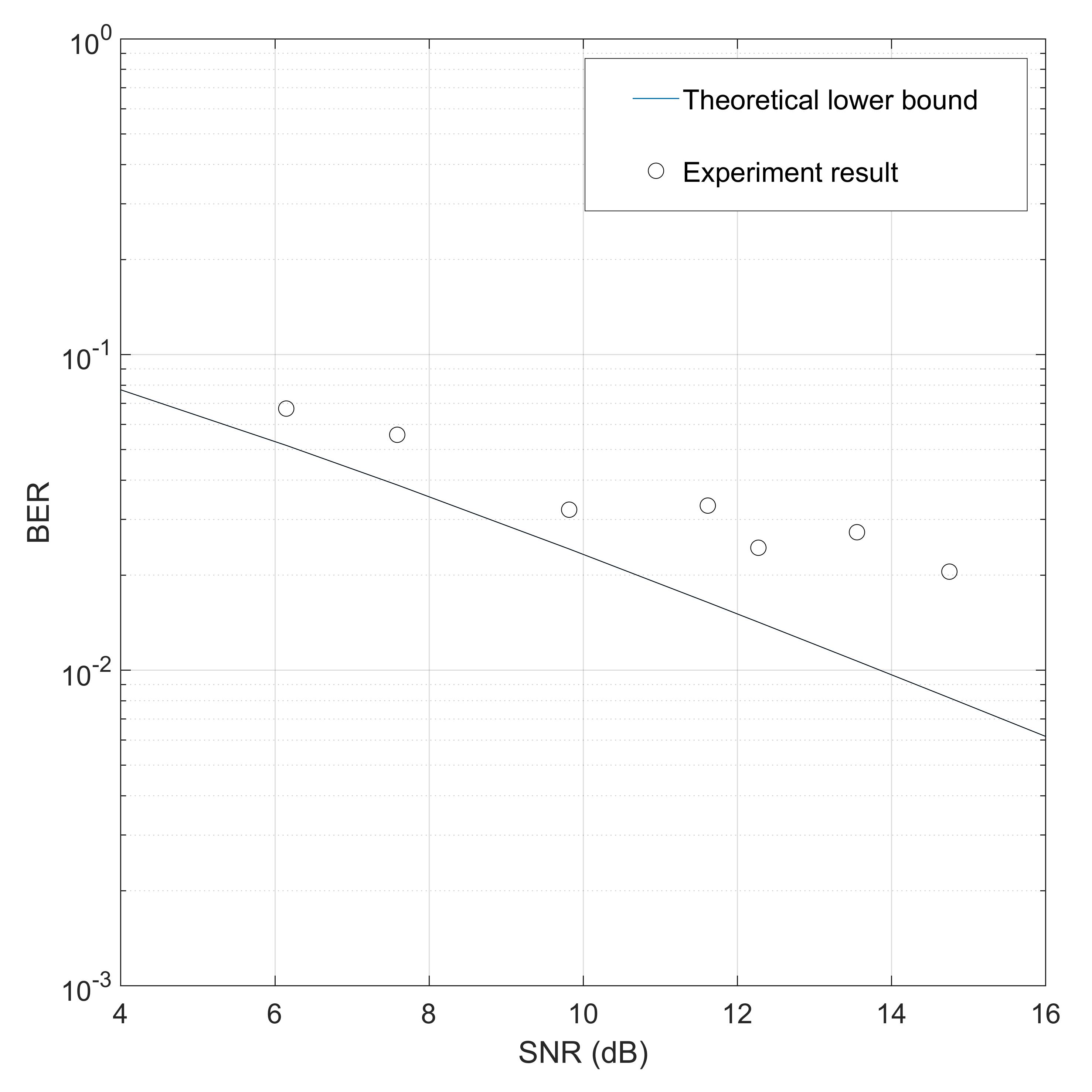}
\caption{BER vs SNR of experiment and theoretical results. The degree of the LDPC code is 128.}
\label{BER:pic1}
\end{figure}

\begin{figure}[!ht]
\centering
\includegraphics[width=8cm,height = 6cm]{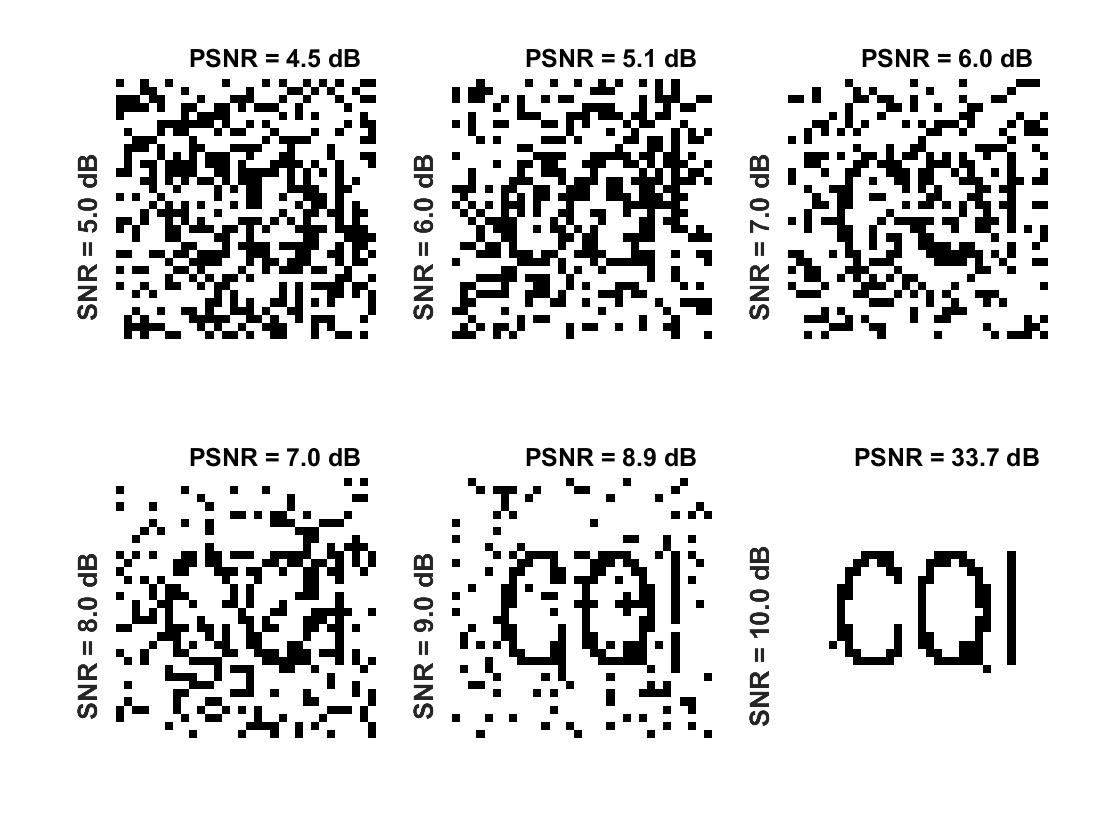}
\caption{The recovered image for different SNR values. The degree of the LDPC code is 128.}
\label{PSNR_vs_SNR:pic3}
\end{figure}

\begin{figure}[!ht]
\centering
\includegraphics[width=8cm,height = 8cm]{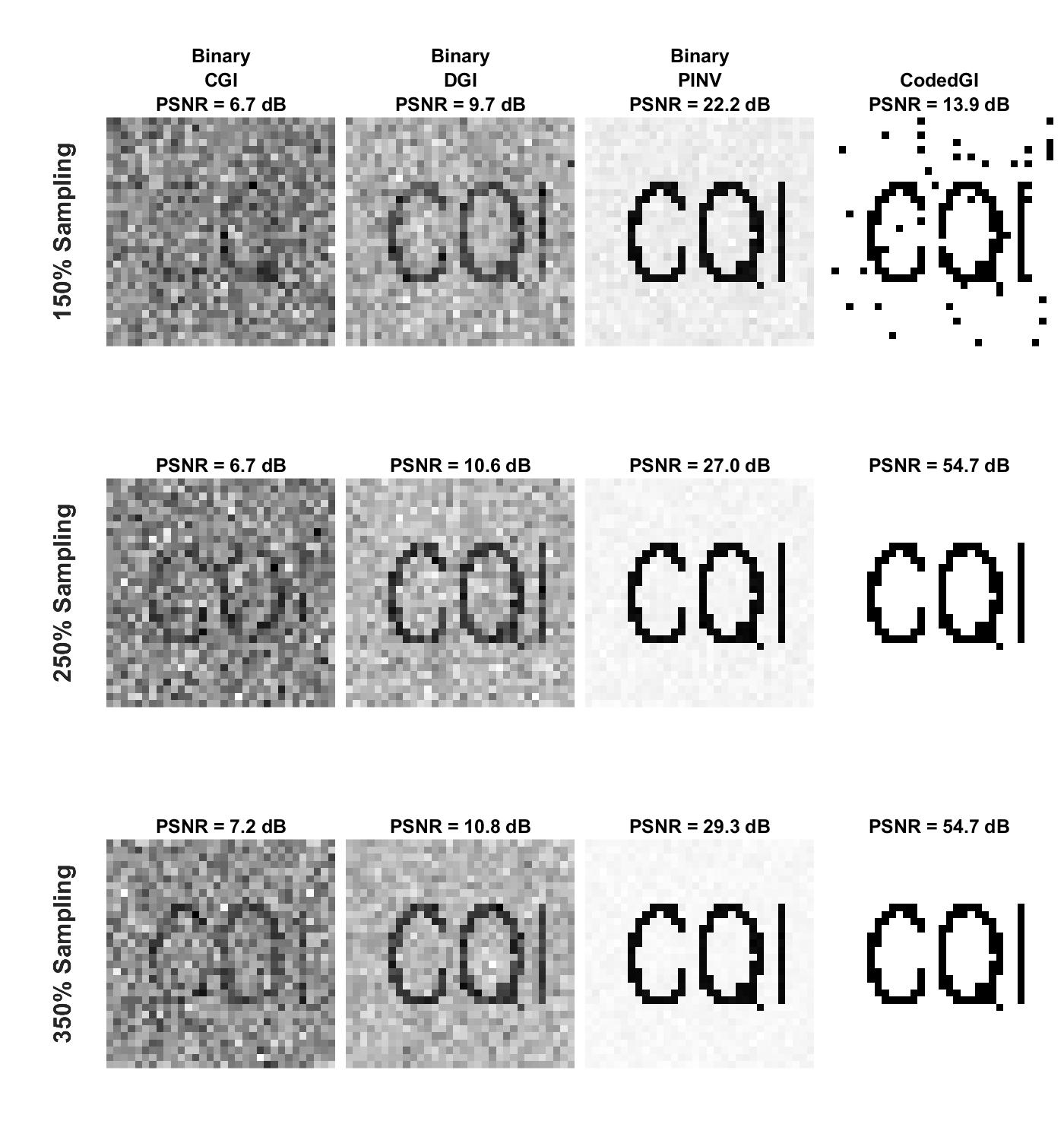}
\caption{Random Binary field illumination vs. LDPC-coded GI when SNR=10 dB. The degree of the LDPC code is 128.}
\label{PSNR:bin}
\end{figure}

\begin{figure*}[!th]
\centering
\includegraphics[width=1.8\columnwidth,height=0.46\linewidth]{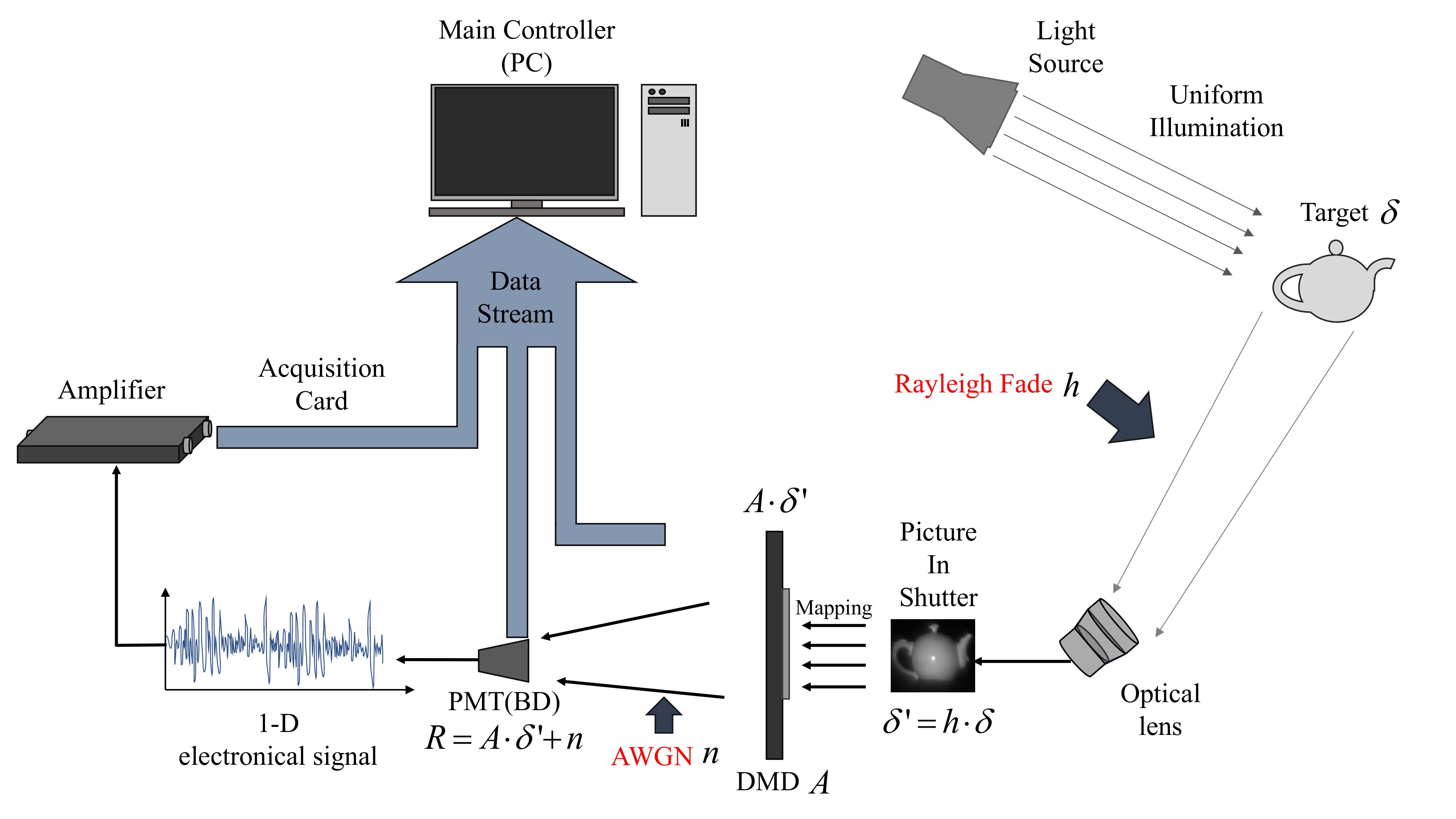}
\caption{Schematic diagram of the experiment platform. }
\label{shiyi}
\end{figure*}

\begin{figure*}[!h]
\centering
\includegraphics[width=2.0\columnwidth,height=0.5\linewidth]{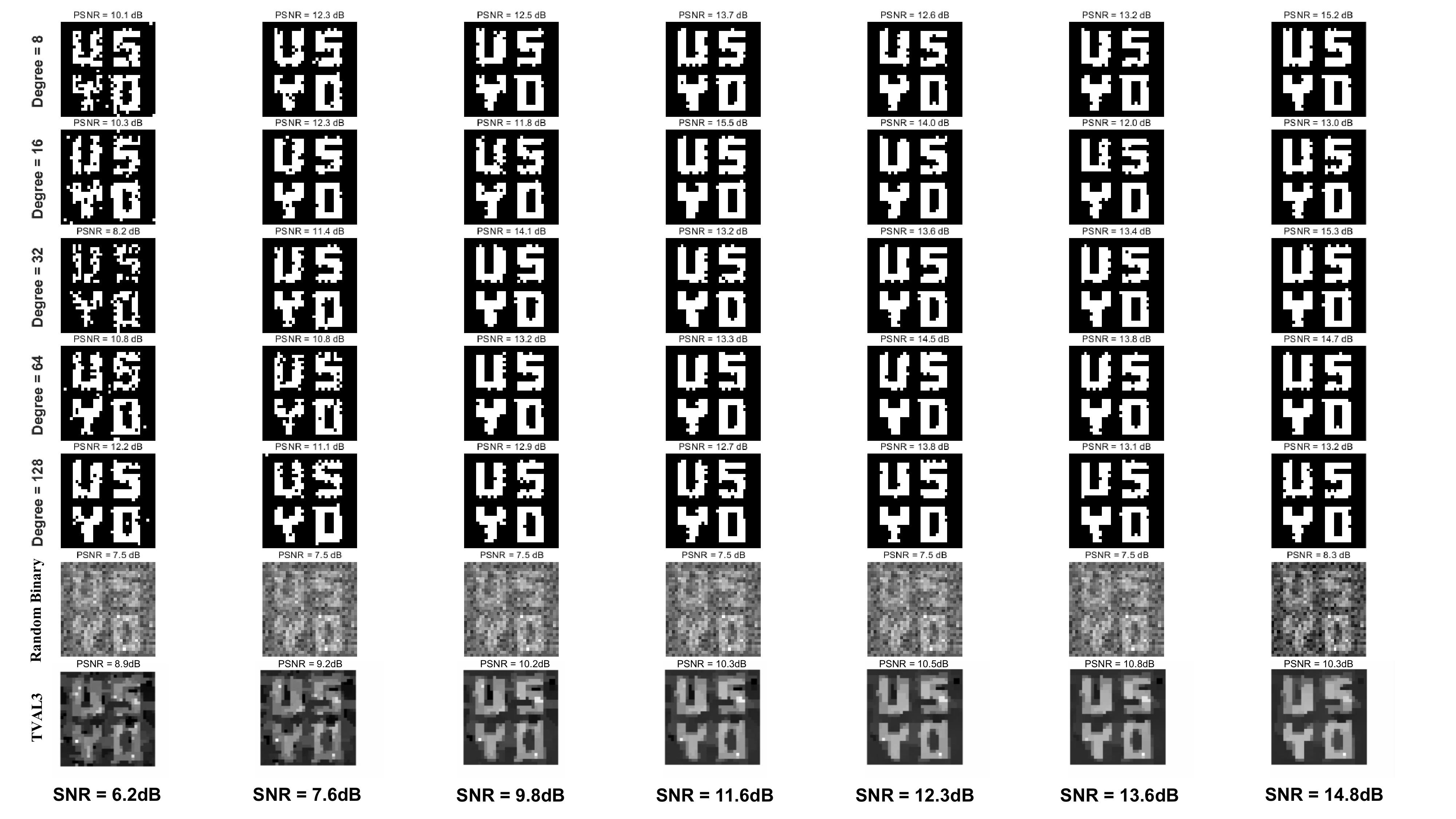}
\caption{The experiment results of the LDPC-coded GI algorithm under different conditions. The x-axis shows that the illumination intensity increases gradually from left to right, and the y-axis shows how the change of matrix sparseness of illuminating patterns influences the imaging result. In the last row, a set of the results using a conventional random binary field illumination correlation algorithm is listed for comparison. It should be emphasized that all the experiment results are obtained under 32 times sampling conditions.}
\label{duibi3}
\end{figure*}

A $[N, M, \Omega_{\omega(\textbf{p})}]$ LDPC code is employed to encode the whole field of view, and a group of random binary codes is set as a comparison. We divide the target imaging plane where an object is located into 32 by 32 pixels, which means the edge length of the LDPC code is $M=1024$. We consider the encoding process using a systematic LDPC code with regular degree distribution. For degree $8$ as an example, i.e., $\Omega_{\omega(\textbf{p})}=x^8$, each illumination randomly selected 8 pixels and combined to generate one output symbol. After we performed $N$ times of illuminations, an image of the target can be obtained by using the BP decoding algorithm. We set $N$ equal to $2048$, i.e., $2$ times of the number of pixels $M$. To calculate the peak signal-to-noise ratio (PSNR), we normalized the original and recovered images so that their values were in the interval $[0,1]$. The BER and PSNR results for each $N$ value are obtained by averaging the results of $10$ experiments.

\begin{figure*}[!ht]
\centering
\includegraphics[width=17cm,height = 3.2cm]{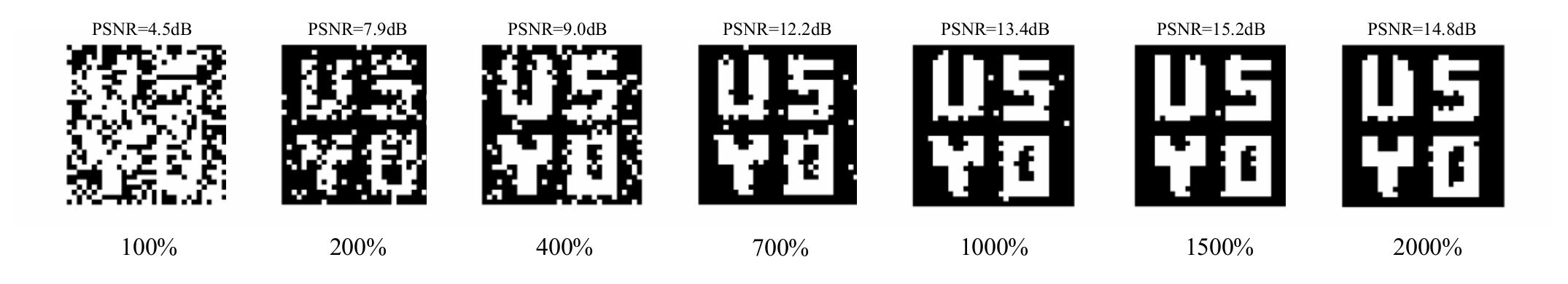}
\caption{The reconstruction results obtained under different sampling rates. The degree of the LDPC code is 128, and the irradiation SNR of the illuminating source is 14.8 dB.}
\label{sample_rate}
\end{figure*}

\begin{figure}[!t]
\centering
\includegraphics[width=0.35\columnwidth,height=0.3\linewidth]{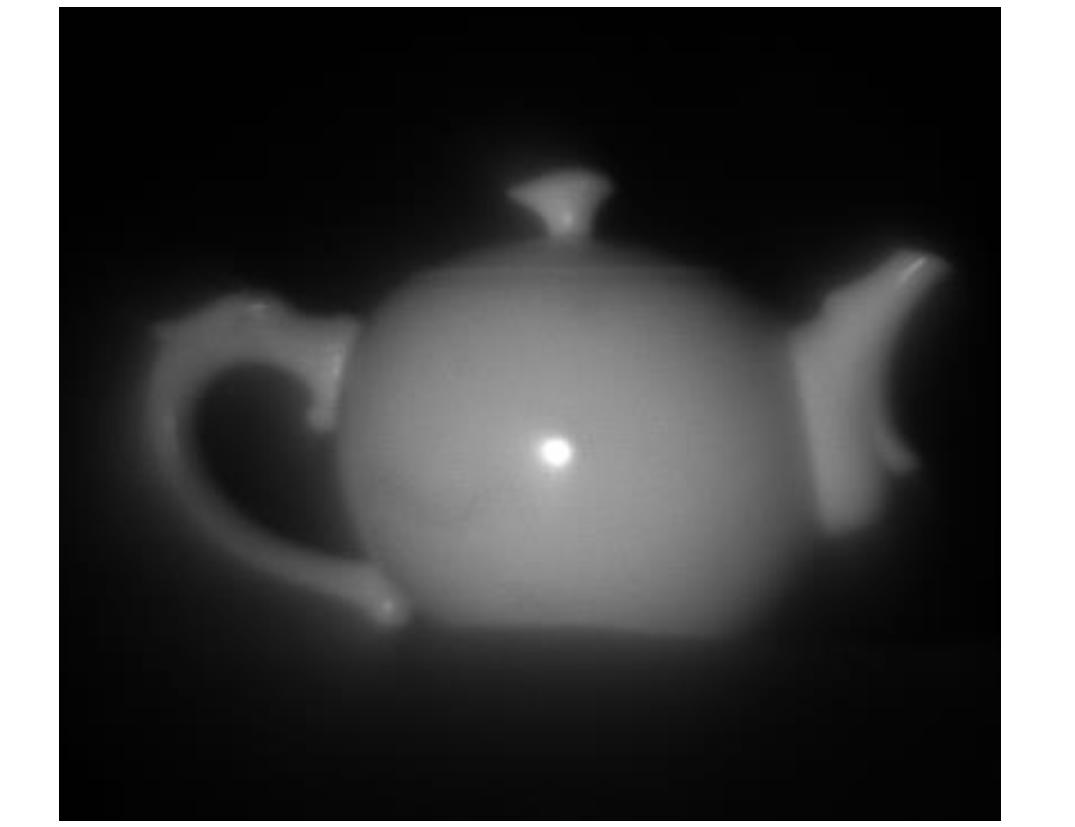}
\caption{A physical image of the teapot in the dark taken by the camera shutter at a certain point. The main body of this teapot is a sphere with a strong reflection point in the center. Such an object with a 3-D depth is considered suitable for testing grayscale imaging performance. }

\label{teapot}
\end{figure}

\begin{figure}[!t]
\centering
\includegraphics[width=1\columnwidth,height=0.6\linewidth]{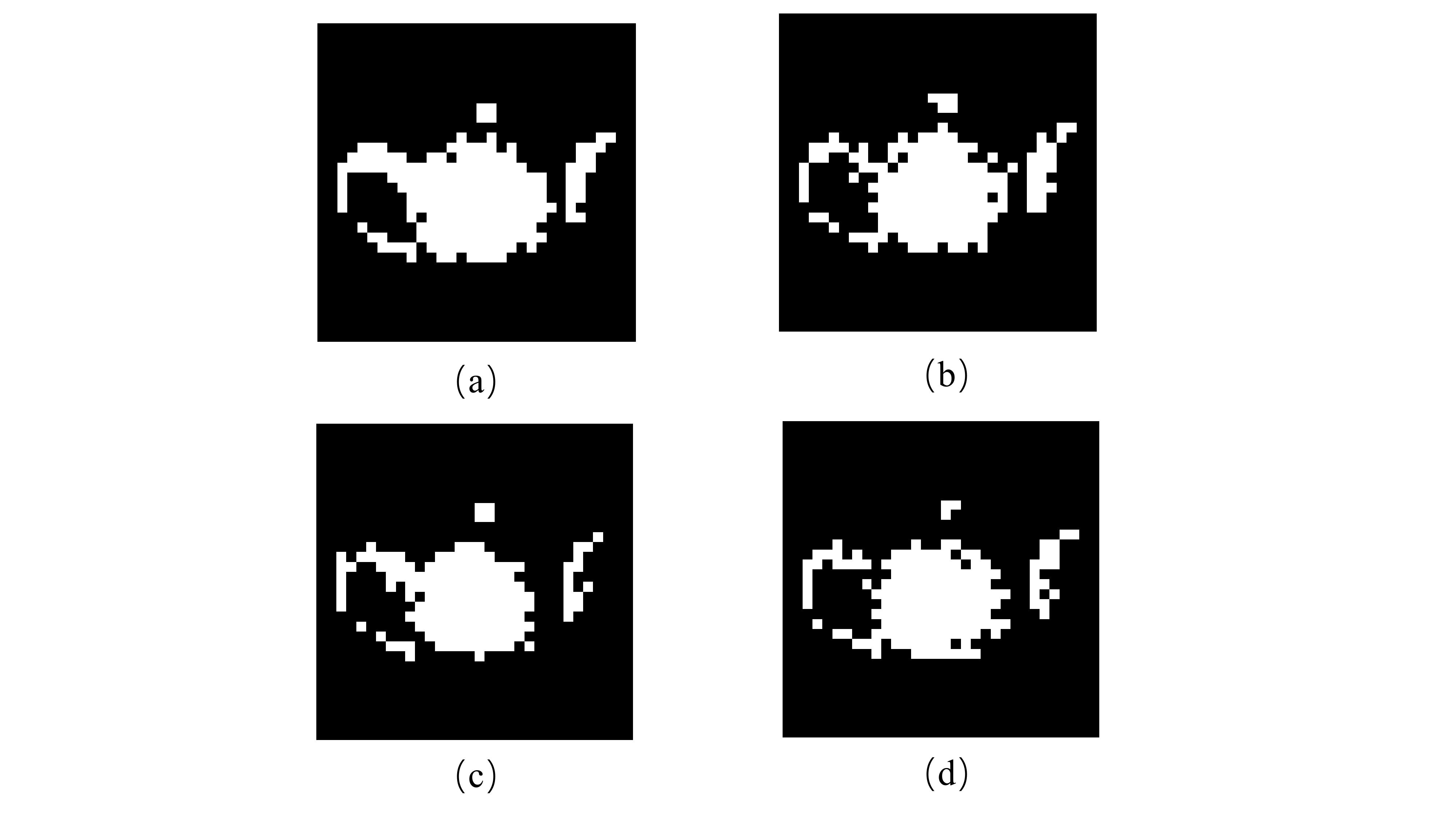}
\caption{The reconstruction results of the non-binary target using the LDPC-based coded GI. The degree distributions of the LDPC code are (a) 128; (b) 64; (c) 32; and (d) 16, respectively. All the experiments are finished under 32 times sampling and the same uniform illuminating light source whose power is $100W$.}
\label{teapot2}
\end{figure}

Fig. \ref{BER:pic1} shows the experiment result and the theoretical lower bound of BER versus signal-to-noise ratio (SNR) for the LDPC-coded GI, in which the degree of the LDPC code is 128. The experiment result and the theoretical result are plotted in dots and lines, respectively. We observe that the experiment result is ideally lower bound by the derived theoretical BER lower bound, and the BER decreases with the increasing of SNR.

Fig. \ref{PSNR_vs_SNR:pic3} shows the recovered image for different SNR values. From the figure, we observe that a recognizable image can only be obtained at a relatively high SNR condition with only twice the sampling rate. When the SNR is no greater than 6 dB, the noise makes the target almost completely inundated.

We also compared the proposed LDPC-coded GI with traditional GI schemes, such as computational GI (CGI), differential GI (DGI)\cite{DGI}, and pseudo-inverse GI (PINV)\cite{PINV,PINV2}. As shown in Fig. \ref{PSNR:bin}, the results obtained from conventional GI schemes are more or less interfered with by the background noise, but the LDPC-coded GI can obtain a perfect binary image that is unaffected by background noise.

\section{EXPERIMENT RESULTS}
\label{6}
\subsection{Experiment Setup}

This section describes the experiment results of the LDPC-coded GI for binary and three-dimensional (3-D) targets. Fig. \ref{shiyi} shows the imaging experiment setup. In this experiment, the channel coding is used to overcome the distortion that occurred in the DMD. A passive optical system is adopted. We set a binary target consisting of four letters "\textbf{USYD}" within the field of view. The distance between the target plane and the receiver is $1.6$m. A uniform artificial light source is used to illuminate the target plane. The target information $\delta$ has suffered Rayleigh fade during transmission. After the image is modulated by the back-end LDPC code, a 1-D signal is received by the bucket detector. The reflected image of the field of view received by the lens is mapped into the DMD. After the back-end modulation of the DMD, the modulated image energy is collected by the photomultiplier (PMT), which is a bucket detector functionally. \\

To keep consistent with the simulation configurations, the target plane is divided into 32 by 32 sub-planes, equivalent to transmitting the target imaging information through 1024 channels. Considering several degree distributions leads to duty ratios varying, whose values are shown in Table.~\ref{table:1}, even under the same illuminating conditions, the detecting SNR of echo signals with different degree distributions are different from each other. Thus, the irradiation SNR on the target plane is adopted as the parameter to measure different source illumination.

\begin{figure*}[!ht]
\centering
\includegraphics[width=2\columnwidth,height=0.55\linewidth]{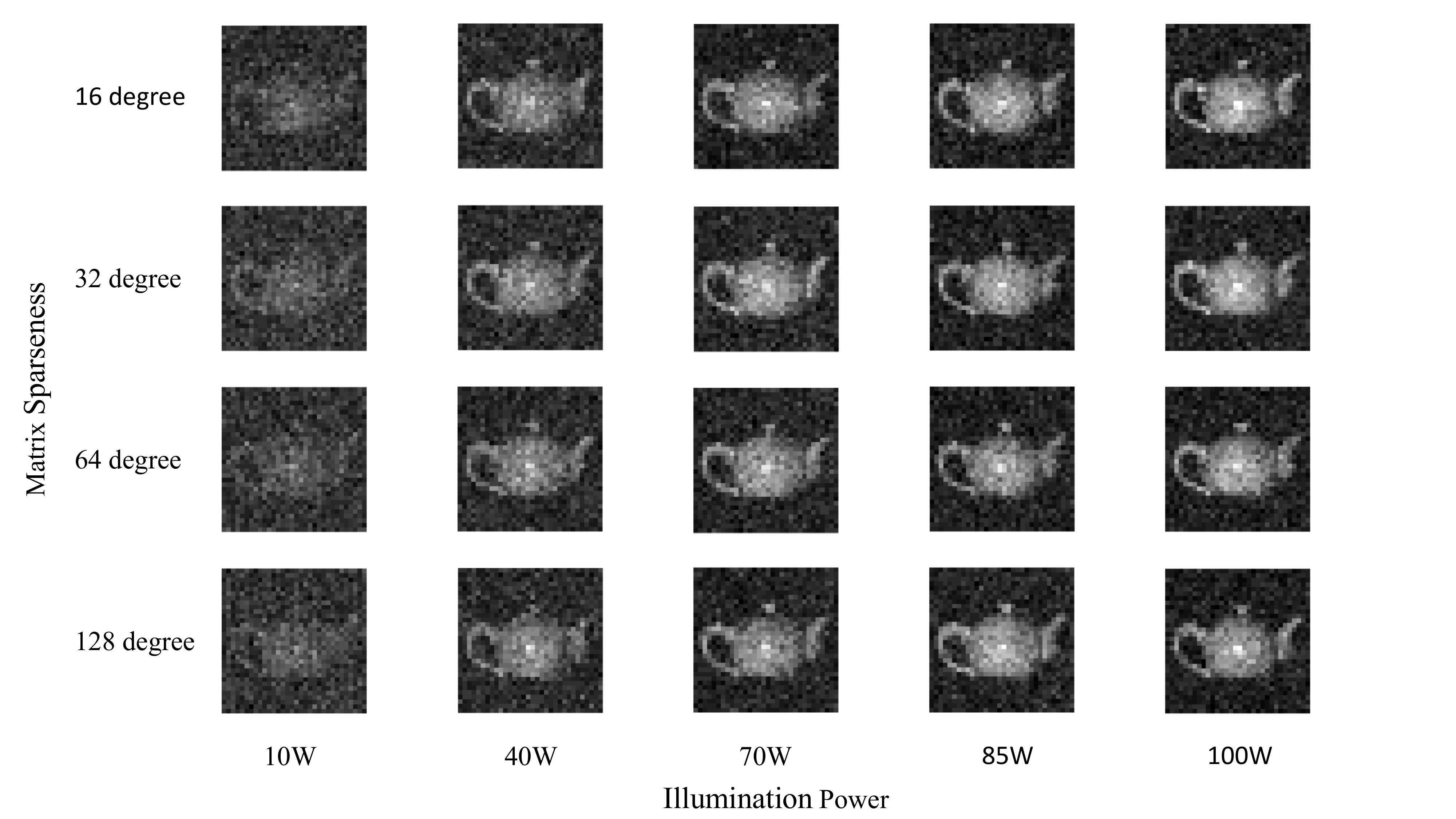}
\caption{The reconstruction results obtained under different sampling rates. Since the target surface is a sphere, the radiation SNR of the target surface is difficult to calibrate. We use the power of the light source as a reference.}
\label{gray}
\end{figure*}

\subsection{Binary Target Reconstruction}
In Fig. \ref{duibi3}, the reconstructed results towards a binary target by the proposed algorithm have been given under the different illuminating intensities and speckle field organizing methods. All results are reconstructed under a 32 times sampling rate. The imaging quality is increased with the irradiation SNR for the three reconstruction algorithms. And the AWGN noise model has been reconstructed as demonstrated in Section.\ref{4}. According to the fading channel model, it is noticed that the binary target reconstructed by LDPC-coded GI is more robust against noise. \\

\begin{table}[!t]
\centering
\caption{Duty ratio of different degree distribution.}
 \begin{tabular}{|p{4cm} | p{2cm} |} 
 \hline
 Coding methods & Duty Ratio\\ 
 \hline
 LDPC 8 degree & 0.78\% \\ 
 \hline
 LDPC 16 degree & 1.56\% \\
\hline
 LDPC 32 degree & 3.13\% \\
\hline
LDPC 64 degree & 6.25\% \\
 \hline
LDPC 128 degree & 12.5\% \\
\hline
 Random Speckle & 15\% \\
\hline
 \end{tabular}
 \label{table:1}
\end{table}

To test the effect of sampling rate on the imaging quality, a series of reconstructed results under different sampling rates are obtained in Fig. \ref{sample_rate}. The experiment results verify the correctness of the simulation results. The target image can barely be reconstructed under system thermal and background noise interference in one-time sampling. When the sampling rate exceeds $700\%$, the noise is almost filtered out from the reconstructed images.\\

\subsection{3-D Target Reconstruction}
Considering there are few ideal binary targets exist in realistic detecting scenarios, and a natural scene target has a 3-D depth of the field of view. The binary image cannot provide information about the object with an uneven surface. Still, the ability of the imaging algorithm to describe these uneven surfaces is often the key to target recognition. The teapot shown in Fig. \ref{teapot} is a classic non-binary object whose main body is a sphere. The reflectance around the teapot is too low for the hard decision probability so these areas are often judged as noise, as shown in Fig. \ref{teapot2}. The randomness of each result will lead to the inability to obtain the exact target contour and size.\\

To acquire the 3-D target image, we attempted to reconstruct the gray-scale image based on the LDPC-coded GI. Since the whole reconstruction process is achieved in the probability domain, the single hard decisions threshold cannot distinguish different reflectance regions. Therefore, if the detection process is sampled multiple times and the data sampled for each time is evaluated as a binary image. The mean value of these binary images would theoretically be the gray image of the target when the sample rate is large enough. Assuming that we need a grayscale image of $n$ bits, then $N=2^{n}$ sampled time is required to reconstruct the target image. If the $N$-th sampling data reconstruction image is $P_{N}(x,y)$, the final grayscale image can be expressed as
\begin{equation}
P(x,y) = 1/N \sum_{N=1}^{2^n} P_{N}(x,y),
\end{equation}
where $P(x,y)$ is the final grayscale image. \\

Limited by the PMT's efficiency, only 32 information cycles can be recorded during a single detection. Thus, this experiment can only test the reconstruction effect of a 5-bit grayscale image. A teapot, placed 5 meters from the radar platform, is illuminated with a uniform white light source in the dark environment. Since the main body of the teapot is spherical, there must be a total reflection spot, and there is a gradual decline in reflectivity near this 'light spot.' It is observed that the reconstruction quality is distinct enough for the target recognition. The 3-D spherical reflectance distribution is described clearly when the illuminating power is greater than 40W. The feasibility of the LDPC-coded GI for grayscale image reconstruction is verified preliminary in this series of experiments as shown in Fig. \ref{gray}. Besides, the BP decoding algorithm is designed based on the channel coding theorem, and its object aims at binary information reconstruction. The hard decision threshold should be re-designed for a multiple interval distribution according to the probability domain information to obtain more accurate and high-bit grayscale images for imaging systems.

\section{CONCLUSION}
\label{7}
We proposed a linear block code-assisted computational GI scheme based on LDPC code and analyzed the theoretical imaging performance of the proposed system. This is the first effort to analytically point out that LDPC-coded computational GI can eliminate the distortion and interference in imaging processing. Meanwhile, the recursive belief propagation will not bring extra complexity to the imaging system. A series of numerical simulations and experiments were used to confirm the viability of the LDPC-coded GI. It's worth noticing that this passive detecting system can only eliminate noise interference, due to the propagation model is not established. For the interference generated by the propagation, we have two schemes to overcome it in the following work: one is to establish the corresponding channel fading model, and use the other channel code to eliminate it; the other is to adopt coherent detection, which coherently eliminates the interference by the coherence of the round-trip path of the propagation. This work also shows some attempts at LDPC-coded GI-based grayscale picture reconstruction in the temporal dimension, demonstrating the necessity for additional studies to comprehend the GI's proposed grayscale image reconstruction technique properly.  When the fading model of scattering medium is determined, the distortion caused by the scattering medium can be completely eliminated, thus the channel-coded GI can be applied in underwater imaging, multi-scattering medium imaging or medical imaging. Furthermore, all kinds of high dimensional information in the light field, such as reflectivity, phase, frequency, and etc., can be encoded by various codes, thus the optical encryption\cite{EncryptedGI} can also be applied.

\ifCLASSOPTIONcaptionsoff
  \newpage
\fi



\end{document}